\documentclass[float,twocolumn,aps,prl]{revtex4}
\usepackage{graphicx}
\usepackage{booktabs}
\usepackage{amsfonts, amsmath, amsthm, amssymb,mathrsfs} % For math f
\usepackage{mathtools}
\usepackage{wrapfig, tikz,url,lipsum,float}
\usepackage[normalem]{ulem}
\usepackage{bm}
\usepackage{subfigure}
\usepackage[colorlinks=true,linkcolor=blue,citecolor=blue]{hyperref}
\usepackage{hyperref}
\usepackage{footnote}
\usepackage{cleveref}
\usepackage[utf8]{inputenc}
\usepackage{xcolor}
\usepackage{natbib}
\usepackage{ulem}
\usepackage{soul}
\label{Off-Spec_Kinematics}

\begin{document}

\date{Revised \today}
\title{\bf Clustering of lipids driven by integrin }

\author{Tapas Singha} 
\affiliation{Institut Curie, PSL Research University, CNRS, UMR 168, Paris, France}
\affiliation{Department of Polymer Science and Engineering,
University of Massachusetts, Amherst, Massachusetts 01003, USA}

\author{Anirban Polley}
\affiliation{Shanmugha Arts, Science, Technology and Research Academy, Tirumalaisamudram, Thanjavur, Tamilnadu 613401, India} 
\affiliation{National Centre for Biological Sciences, UAS-GKVK Campus, Bellary Road, Bangalore 560065, India}
\author{Mustansir Barma}
\affiliation{TIFR Centre for Interdisciplinary Sciences, Tata Institute of Fundamental Research, Gopanpally, Hyderabad 500107, India}

\begin{abstract}Integrin is an important transmembrane receptor protein which remodels the actin network 
and anchors the cell membrane towards the extracellular matrix via mechanochemical pathways.  
The clustering of specific lipids and lipid-anchored proteins, which is essential for a certain type of endocytosis process, is facilitated at integrin-mediated active regions. To study this, we propose a  minimal exactly solvable model which includes the interplay of stochastic shuttling between 
integrin on and off states with the intrinsic dynamics of the membrane. We  obtain an analytic expression for the deformation and local membrane velocity, and thereby the evolution of clustering mediated by a single integrin. The deformation, velocity and lipid clustering evolve nonmonotonically and their dependences on the stochastic shuttling timescales and membrane properties are elucidated.
\end{abstract}

\maketitle
\section{I. Introduction}
The smooth functioning and regulation of the cell involves the meshing of numerous mechano-chemical pathways. These pathways and their intersections often involve multi-tasking components that fulfil multiple functions, which however are not always evident at first sight. The identification of multi-tasking components and their functions remains an important open problem in cell biology.
 
 A prime example of a multitasker is integrin, a bidirectional signaling receptor protein found in the plasma membrane of the cell. Integrin mechanically controls many aspects of cell life, including adhesion, migration, proliferation and  differentiation \cite{Hynes2002673,Paulina2014}. It also participates in transmitting information about the chemical and physical state of the ligands into the cell (out-to-in) to facilitate cell-migration, cell-survival and growth, and, separately, takes part in transmitting the signals outside (in-to-out) to regulate cell adhesion \cite{Campbell2011}. In recent experimental studies, another important, quite different function of integrin is recognized  namely its role in signaling cascade of sphingomyelin and cholesterol and nano-clustering of glycosylphosphatidylinositol anchored proteins (GPI-APs)\cite{Kalappurakkal2019,Raghupathy_2015,Simons1997}, which is essential for GPI-APs mediated endocytosis \cite{Sharma2004577,Doherty2009}. In this work, we discuss the underlying mechanism of the GPI-APs clustering via a minimal exactly solvable model. As we will argue below, integrin accomplishes the clustering through a two-step process, which involves the actomyosin network in the cell membrane as an intermediary.

Several theoretical and experimental studies suggest that actin-membrane composites play a significant role in cluster formation on
the plasma membrane (PM) \cite{Pierre2011, Das2016a}. In both leaflets of the PM, there is an inherent asymmetry with respect to both the composition and numbers of protein and lipids \cite{Sharma2004577,Raghupathy_2015}. We classify PM components which are able to regulate the actin machinery as active and  those which are controlled by the actin machinery as passive. Active molecules like integrin can interact with actin and regulate local actin dynamics \cite{Hynes2002673}, thereby influencing the organization of the passive particles. Passive particles such as, GPI-AP, phosphatidylserine (PS) lipids interact with organized actin filaments, and so fluctuations of the actin density lead to clustering of lipids as described below. Finally, there are also inert molecules such as short unsaturated acyl chain lipids which do not interact with actin at all, in either an active or passive sense \cite{Raghupathy_2015}.

How does the actin network couple to the cell membrane, and how do actin and passive-lipid interactions arise? The actin filaments of the actomyosin network move stochastically at the cell cortex, and their collective dynamics causes density fluctuations which generate localised contractile stress and radial arrangement of filaments i.e., asters \cite{Das2016a}. This contractile stress affects the PM through the actin filaments and in addition, polymerization of actin filaments also generates force on the membrane \cite{Chhabra2007}. As the cortical actin network is closely juxtaposed to the PM, the contribution of the stress along the direction normal to the membrane changes the local curvature of the PM. On the other hand, the organized filaments oriented towards the deformation create a zone into which PS lipids at the inner leaflet get advected \cite{Raghupathy_2015}. Further, PS lipids interact across the bilayer with long-saturated-acyl-chain-containing GPI-APs at the outer leaflet, thereby facilitating GPI-AP nanoclustering \cite{Raghupathy_2015}.

Considering a zero average stochastic force arising from actin polymerization, an earlier theoretical study explored the different regimes when lipid density is driven by membrane fluctuations \cite{Pierre2011}. Though several studies \cite{Edidin2006,Goswami2008} suggest that membrane components interact with the organized actomyosin, and form a cluster, the mechanism remains elusive. A recent experiment reveals \cite{Kalappurakkal2019} that upon binding to the ligand, integrin activates a signaling cascade that triggers actomyosin nucleation via specific proteins like formins, talin, and vinculin. In this work, we argue that integrin together with the actin filaments induces the clustering of GPI-APs and PS lipids.

Integrins are heterodimer transmembrane poteins consisting of $\alpha$ and $\beta$ subunits, which link the extracellular matrix (ECM) to the cytoskeleton. Different varieties of $\alpha$ and $\beta$ subunits assemble into 24 types of integrin with different binding properties \citep{Campbell2011}. An important point is that integrin exists in two different states with different conformations, namely an active on-state, and a passive off-state \cite{Campbell2011,Bidone2019}.  

\begin{figure}
%\hspace{-1.60cm}
\begin{center}
        \includegraphics[width=0.90\linewidth, height=0.12\textheight]{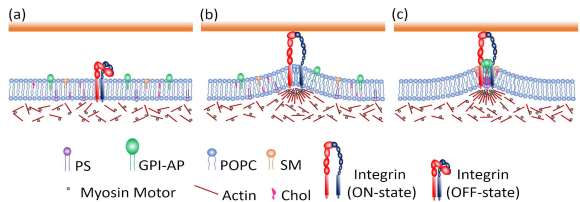} 
\end{center}        
\caption{Schematic of the two-step mechanism of specific lipid clustering regulated by integrin.
(a) shows the off-state integrin (red and black denote its two subunits of the dimer
configuration) embedded in a plasma membrane (PM) comprising POPC (blue), SM (orange) and Chol (pink)
with upper leaflet GPI-AP (green) and inner leaflet PS (purple) while actin filaments (red straight lines) and myosin motor (black) reside beneath the PM. The extracellular-matrix is shown by thick orange line near the PM. (b) shows the
on-state integrin adhering with the ECM, resulting in reorganizing the actin network and creating a local deformation of the PM. (c) shows the clustering of specific lipids such as upper leaflet GPI-AP and
inner leaflet PS associated with SM and Chol initiated by the adhesion of on-state integrin with ECM to PM.}
\label{actin}
\end{figure}

In-to-out and out-to-in signaling by active integrin with the help of other proteins results in the formation of compact and complex assemblies called focal adhesions (FAs), which facilitate the anchoring of the surrounding ECM to the actin-cytoskeleton. Integrins stochastically switch between on and off states rather than staying in a long-lived on state \cite{Rossier2012}. The on-off dynamics of the integrin and the membrane relaxation 
collectively regulate the formation of the dynamic nano-scale organization.  Further, a few auxiliary
proteins such as fibronectin, actin-binding protein talin and vinculin bind to the integrin and immobilize them at FAs. By contrast, outside the FAs, integrins are found to exhibit high diffusion and talin cannot be sustainably bound to the diffusing integrin \cite{Ivaska2012}.

\emph{Mechanisms:} In order to understand the mechanism by which integrin at FAs is able to organize the clustering of specific lipids (GPI-APs, PS), we propose a new two-step mechanism via a minimal exactly solvable model in two dimensions. Here we study only the out-to-in signaling of a single integrin which binds to the ECM and anchors the PM.  
   
 Customarily, it is envisaged that the lipids are driven by the collective stochastic fluctuations of the actin filaments and myosin motors at the cortex resulting in aster-like arrangement of actin filaments \cite{Das2016a}. Here we propose a new mechanism in which on-state integrin plays a crucial role: first, it reorganizes the randomly oriented actin network around the  location of the integrin; and secondly, it anchors the membrane towards ECM, and creates a deformation around which the actin filaments and myosin motors reorganize themselves to maintain the compactness of the actomyosin composite. In other words, the integrin on the one hand chemically reorganizes the actin network and on the other hand, it mechanically creates space to pack the filaments better and enhance the polarisation of the filaments. Both processes help lipid clustering. Integrin-induced organization of the actin filaments spreads till the on-state integrin (ONI) goes to the off-state. As both the processes, namely reorganization of actin-filaments, lipids and GPI-APs, and   deformation of the membrane occur simultaneously mediated by the integrin, we assume the second process provides an indirect way of measuring the first (estimating lipids and GPI-APs clustering).

In short, we consider a single integrin on a two-dimensional  membrane, which anchors the membrane in its on-state, and diffuses in the off state. The deformation of the membrane facilitates the organization of polarized actin filaments into asters. Since the lipids get advected by these asters, \emph{integrin-induced deformation} leading to asters formation facilitates the lipid and GPI-APs clustering on the cell membrane.

As the diffusivity of the lighter lipids is much higher ($\sim 10$-fold) compared to that of integrin \cite{Cheng2020,Saha2015}, the lipids and GPI-APs are expected to follow the induced average deformation, extremely fast. We take the process to occur instantaneously. For example, the diffusion constant of GPI-APs are $\sim 2.5 \mu m^2/sec$ \cite{Saha2015} whereas the diffusion constant of off-state integrin is approximately $ \sim 0.3 \mu m^2/sec$\cite{Arora2012}. Furthermore, a typical size of an aster is $30-450\,nm$, and typical on-time of the integrin is $3$ to $33\, sec$, therefore we consider that the timescale of GPI-APs movement on the PM is much faster compared to the integrin switch times and the lifetime of the associated asters. Note that other membrane proteins, that do not actively remodel the actin-network, would not lead to clustering of PS lipids and GPI-APs, though they may make membrane deformations.

\section{II. Model and Results}
In the absence of coupling to the integrin, the free energy of the cell membrane is taken to have the form \cite{ Helfrich1973,Jane2019}  
\begin{align}
F[h] = \int d^2 \mathbf{r} \left \{ \frac{\nu}{2} (\nabla h)^2 + \frac{\kappa}{2} (\nabla^2 h)^2 + \frac{\gamma}{2}[h(\mathbf{r},t)]^2 \right\} 
\label{HelfrichPot}    
\end{align}
where $h(\mathbf{r},t)$ is a deformation in membrane height in two dimensions at time $t$. The three terms on the right-hand side represent respectively the elastic energy, the bending energy, and a harmonic potential which may capture the effect of steric interaction, adhesion of the cortical actomyosin to the membrane, or binding of the glyco-protein to ECM. It inhibits large-scale deformation so that the overall shape and size of the cell is maintained.

In the on-state, integrin remains stationary and anchors the PM towards the ECM whereas in its off-state it detaches and diffuses. The transitions from on-to-off and off-to-on are stochastic, and we may write a Fokker-Planck equation, taking the rates to be independent of the membrane deformation. Using $+$ and $-$ to denote on and off states, respectively and $P_{+}(\mathbf{r},t)$ and $P_{-}(\mathbf{r},t)$ as the corresponding on and off state probabilities, we have
\begin{eqnarray}
\frac{\partial P_{+}(\mathbf{r},t)}{\partial t} &=& -\lambda_{+} P_{+}(\mathbf{r},t) + \lambda_{-} P_{-}(\mathbf{r},t)    \nonumber \\
\frac{\partial P_{-}(\mathbf{r},t)}{\partial t} &=& -\lambda_{-} P_{-} + \lambda_{+} P_{+} + D \, \nabla^2 P_{-}(\mathbf{r},t). \nonumber \\
\label{FKP_Integrin} 
\end{eqnarray}
Here $D$ is the diffusivity in the off-state. As the integrin is much larger in size compared with the other transmembrane proteins \cite{Bidone2019}, we assume that it does not get advected by the curvature of the cell membrane. The on-to-off (off-to-on) rates $\lambda_{+}$ ($\lambda_{-}$) correspond to average lifetimes $\langle \tau_{\pm} \rangle=1/\lambda_{\pm}$.
We use the typical values of $\lambda_{\pm}$ as inputs from ref.\cite{Cheng2020}. Conservation of probability implies that  $\int \, d^2 \mathbf{r} \,\, [P_{+}(\mathbf{r},t)+P_{-}(\mathbf{r},t)]=1$. 

To model the effect of the integrin on the dynamics of the cell membrane, we write
\begin{align}
\frac{\partial h(\mathbf{r},t)}{\partial t} = - \frac{1}{\Gamma} \frac{\delta F[h]}{\delta h(\mathbf{r},t)}  + I_0\,\sigma(\mathbf{r},t) + \eta (\mathbf{r},t)
\label{EWMS_Integrin}
\end{align} The first term $\delta F/\delta h$ includes the relaxation processes discussed above. In the second term $I_0$ is the velocity with which integrin induced actin-integrin-ECM complex pulls the PM locally. We roughly estimate the value of $I_0=0.2 \mu m/sec$, using $I_0 = f^{\text{ext}}/\Gamma$ and taking the values  $f^{\text{ext}}\simeq 20pN$, and $\Gamma = 100 pN sec/\mu m$. In Eq.\,(\ref{EWMS_Integrin}), $\sigma(\mathbf{r},t)$ can take two values $\sigma_{+}=1$ (on-state), and $\sigma_{-}=0$ (off-state). The last term  $\eta (\mathbf{r},t)$ incorporates rapid thermal fluctuations which satisfies $\langle \eta\rangle=0$, and $\langle \eta(\mathbf{r},t) \eta(\mathbf{r}',t)\rangle = (4 k_B T/\Gamma) \delta (\mathbf{r}-\mathbf{r'}) \delta(t-t')$. 

Fluctuating mean-zero forces of the type considered in \cite{Pierre2011} and exemplified by the $\eta(\mathbf{r},t)$ term in Eq.\,(\ref{EWMS_Integrin}) have a much smaller effect on the membrane profile than the slower stochastic fluctuations of on-off integrin states. The fact that the latter stochastic sequence has a nonzero average value leads to the formation of a nontrivial mean profile. Therefore, we define the average height field as $H(\mathbf{r},t)=\langle h(\mathbf{r},t) \rangle_{\eta}$, and $\langle \sigma(\mathbf{r},t) \rangle_{\eta}= P_{+}(\mathbf{r},t)$, which satisfy the evolution equation
\begin{equation}
\frac{\partial H(\mathbf{r},t)}{\partial t} = \nu \nabla^2 H  - \kappa \nabla^4 H  -\gamma H + I_0 P_{+}(\mathbf{r},t) \label{EW_Integrin}
\end{equation}

In order to obtain the effect of the integrin on the membrane, we first define  $\widetilde{P}_{\pm}(\mathbf{q},s)=\int d^2 \mathbf{r} \,\, e^{i \mathbf{q} \cdot \mathbf{r}} \int^{\infty}_0 dt \,e^{-s\,t} P_{\pm}(\mathbf{r},t)$
and similarly $\widetilde{H}(q,s)$ in terms of $H(\mathbf{r},t)$ where  $\mathbf{q}$ and $s$ are the Fourier and Laplace variables conjugate to real space and time, respectively. We denote the initial on-state probability by $P_{+}(\mathbf{r},0)=p_0\,\delta(\mathbf{r}-\mathbf{r}_0)$, and off-state probability by $P_{-}(\mathbf{r},0)= (1-p_0) \, \delta(\mathbf{r}-\mathbf{r}_0)$. From Eqs.\,(\ref{FKP_Integrin}) and (\ref{EW_Integrin}), we obtain (see Appendix-A)
\begin{equation}
 \widetilde{P}_+(\mathbf{q},s)= \frac{\left(s+D  \mathbf{q}^2\right) p_0 + \lambda _- }{\left(s+\lambda _-+D \mathbf{q}^2\right) \left(s+\lambda_+\right)-\lambda _- \lambda _+}\, e^{i \mathbf{q} \cdot \mathbf{r}_0}
\label{SolFKPEq1}
\end{equation}
and 
\begin{align}
 \widetilde{H}(\mathbf{q},s)=  \frac{I_0 \, (\lambda _- + \left(s+D  \mathbf{q}^2\right) p_0) e^{i \mathbf{q}.\mathbf{r}_0}  }{(s+ \gamma+\nu \mathbf{q}^2 +\kappa \mathbf{q}^4)(s^2 + (\lambda + D \mathbf{q}^2)s+ \lambda_{+}\, D \, \mathbf{q}^2)}\, 
\label{HtBendingRigidity}
\end{align}
where $\mathbf{r_0}$ is the initial location of the integrin on a two-dimensional membrane.

For the PM, the competition between the bending stiffness $\kappa$ and surface tension $\nu$ sets a crossover length-scale $\ell_c \sim \sqrt{\kappa/\nu}$. The typical range of bending rigidity $\kappa$ for the PM is $10$$-$$25$ $k_B T$ \cite{Marsh2006146, Deserno201511}, while $\nu$ is $2.4\,\, $\text{to}$\,\, 10 \times 10^{-3}k_B T/(nm)^2$ \cite{Gauthier2012}, and $\gamma \simeq 3 \times 10^{-7}$ $k_B T/(nm)^4$. These values lead to a crossover length $\ell_c \simeq 64\, nm$. Thus $\nu$ would be relevant only for $\ell$ larger than $\ell_c$. On length scales $\ell < \ell_c$ relevant for lipid clustering, the dynamics of the membrane is largely controlled only by the bending stiffness $\kappa$. 
%When $\nu$ is small, $\ell_c$ is large, and the bending stiffness controls the membrane dynamics for  length scales satisfying $\ell < \ell_{c}$.

Experimental observation reveals that inside the crowded region of FAs, approximately $86\%$ 
of the integrins are either immobile ($69\%$) or confined ($17\%$), and the diffusion constant of the remaining integrins is also much lower than outside the FAs \cite{Rossier2012}. For simplicity, we first study the immobile ($D=0$) integrin in the zero-tension limit $\nu=0$. From Eq.\,\ref{HtBendingRigidity} in Appendix-A, we obtain the exact expression for the deformation at the location of the integrin at 
$\mathbf{r}_0=0$ as
\begin{equation}
H(\mathbf{0},t)  =  \frac{I_0}{\sqrt{64 \kappa}} \left[ \frac{\lambda_{-}}{\lambda} \frac{\text{Erf}[\sqrt{\gamma t}]}{\sqrt{\gamma}} + a\,  e^{-\lambda t} \frac{\text{Erfi}[\sqrt{(\lambda-\gamma)t}]}{\sqrt{(\lambda-\gamma)}}  \right]
\label{AveDefor}
 \end{equation}
where $a=(p_{0}-\frac{\lambda_{-}}{\lambda})$, $\text{Erf}(x)=\frac{2}{\sqrt{\pi}} \int^{x}_0 dy \, e^{-y^2} $, $\text{Erfi}(x)=\text{Erf}[i x]/i$, and $\lambda=\lambda_{+}+\lambda_{-}.$

\begin{figure}
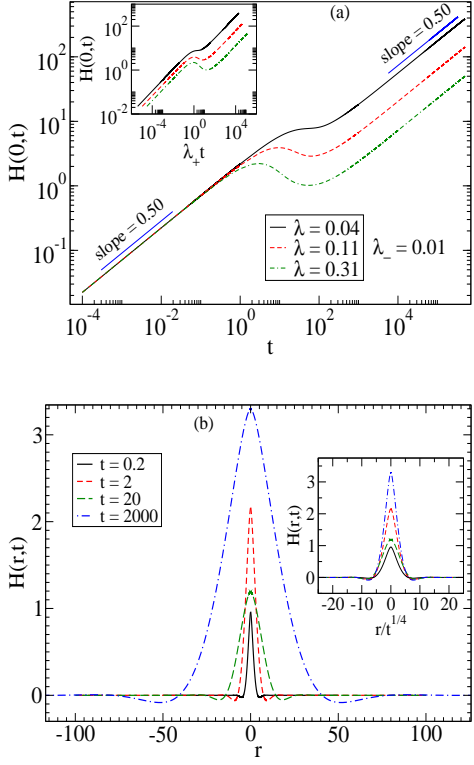

\hspace{-2.0cm}
\centering
\begin{minipage}{0.587\textwidth}
        \includegraphics[width=0.587\linewidth, height=0.20\textheight]{Depth_Vel_Kappa.eps} \\
        \vspace{0.6cm}
        \includegraphics[width=0.587\linewidth, height=0.2\textheight]{HeightProfileArea.eps}
        \end{minipage}%
\caption{Evolution of deformations with time and the deformation profile in space are plotted for a tension-less two-dimensional membrane. (a) The evolution of the deformation at the location of the integrin ($\mathbf{r}_0=0$) is plotted for different on-rates. In the inset, we show how the first maxima of the deformation scales with $\lambda_{+}$. (b) Deformation profile in space is shown where $\mathbf{r}$ is the radial distance from the location of the integrin wherein we choose $\lambda_{-}=0.01\, \text{sec}^{-1}$ and $\lambda =0.31\,\text{sec}^{-1}$ for the deformation profiles \cite{Cheng2020}. For both the plots (a) and (b), we choose $\kappa=10 k_B T$ and $p_0=1.00$}
\label{DeforProf_Depth}
\end{figure}

In the limit of weak harmonic confinement, $\gamma \ll \lambda$, we obtain the simpler form  $
 H(\mathbf{0},t) = \frac{I_0}{\sqrt{16 \pi \kappa}} ( \frac{\lambda_{-}}{\lambda} \sqrt{t}  + \,\, \frac{a}{2} \sqrt{\frac{\pi}{\lambda}} e^{-\lambda t}  \text{Erfi}[\sqrt{\lambda t}] ) $. Inside FAs, the integrins are more often in the on-state, and thus immobile  ($69\pm2 \%$) whereas outside of FAs, this drops to $46 \pm3\%$\cite{Rossier2012}. Therefore, in this study, we take initial on-state probability  $p_{0}>1/2$, leading to
\begin{equation}
 H(\mathbf{0},t)  \simeq \frac{I_0}{\sqrt{16\pi \kappa}} \times
\begin{cases}
 p_{0} \,\, \sqrt{t} & \text{ $\lambda t \ll 1$} \\
 \frac{\lambda_{-}}{\lambda} \,\,  \sqrt{t} & \text{ $ 1 \ll \lambda t \ll \gamma \, t $} \end{cases}
\label{Depth2DKappa}
\end{equation}

\begin{figure}[H]
\begin{minipage}{0.6\textwidth}
        \includegraphics[width=0.6\textwidth, height=0.22\textheight]{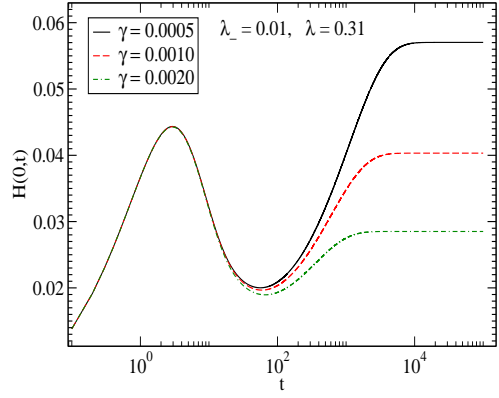}%
\end{minipage}%        
\caption{The plot shows the time evolution of the deformation for different values of $\gamma$ in which we choose $\kappa = 10 \, k_B T$.}
\label{DepthGamma}
\end{figure}

The deformation grows as $t^{1/2}$ in both the small time and long time limits but with different coefficients of $t^{1/2}$ as shown in Fig.\,(\ref{DeforProf_Depth}). Figure\,\ref{DeforProf_Depth} shows the time evolution of deformation profile with $p_{0}=1$. The nonmonotonicity in $H(\mathbf{0},t)$ with $t$ continues to hold as long as $p_0>\lambda_{-}/\lambda$. For large $t$, beyond $\sim 1/\gamma$, the deformation stops increasing and saturates as $\gamma$ provides a brake on the further growth of the deformation. We find that the braking occurs at earlier times with increasing $\gamma$ as shown in Fig. (\ref{DepthGamma}). 

At a given time, the deformation decreases with increasing lateral separation from the location of the integrin as shown in Fig.\,(\ref{DeforProf_Depth}b). A good measure of the spread is given by $R(t)$, the location of the first zero of $H(\mathbf{r},t)$.
This spread grows in time and follows a simple power-law $R(t) \sim t^{1/4}$ as shown in the inset of Fig.\,(\ref{DeforProf_Depth}b). It behaves similarly to the length scale of membrane fluctuations, which follow $\sim t^{1/z}$ where the dynamic exponent $z=4$ (when $\kappa \neq 0$ and $\nu=0$ in Eq.\ref{EW_Integrin}).

%\textcolor{blue}{This may be expected as long as the active anchoring does not affect the membrane relaxation process. Because the static integrin directly affects only at $\mathbf{r}=\mathbf{r}_0$ and the corresponding deformation $H(\mathbf{0},t)$ laterally spreads on the membrane governed by the membrane relaxation mechanism (due to $\kappa$) similar to the passive system.}

\begin{figure}
\hspace*{-0.8cm}
\begin{minipage}{0.6\textwidth}
\centering
        \includegraphics[width=0.6\linewidth, height=0.2\textheight]{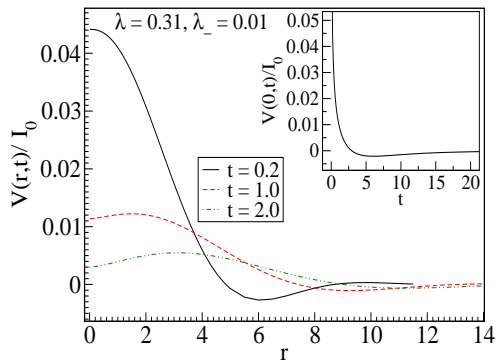}
        \end{minipage}
\caption{Velocity profile of a two-dimensional membrane for different times. Here $r$ is the radial distance from the location of the integrin $r_0=0$. The local velocity of the membrane for different time is shown wherein we choose $p_{0}=1$ and $\kappa=10$. The inset shows how the $V(0,t)$ evolve in time. For the inset, we choose the same value of $\lambda_{-}$, $\lambda$ and $p_0$ as considered for the main curves.}
\label{RateDeformation}
\end{figure}

 We find the anchoring of an ONI deforms the membrane and induces a local velocity along the normal to the membrane at the position of the integrin and its neighborhood for different times as shown in Fig.\,(\ref{RateDeformation}). 
The nonmonotonic variation of the mean-height profile Fig.\,(\ref{DeforProf_Depth}a) implies an interesting behavior of the average velocity $V(\mathbf{r},t)=\partial H/ \partial t$.  
As shown in the inset of Fig.\,(\ref{RateDeformation}), the velocity $V(0,t)$ is high initially and decays as $t^{-1/2}$. It becomes negative as integrin goes to the off state and the membrane relaxes due to $\kappa$, and rises again in the on state of the integrin. We find that $V(0,t)$ decreases faster with increasing $\kappa$ as $V(0,t)\sim\kappa^{-1/2}$ for the two-dimensional membrane. Note that too high a velocity would not be conducive to lipid clustering, as the finite diffusivity of lipids would limit the efficiency of the process. The favorable range of deformation rate depends on many details of the dynamics of the actomyosin network and regulatory proteins, the effects of which lie beyond the scope of this work.  

With these caveats, we turn to the quantification of  integrin-induced clustering on the membrane. To this end, we assume that  there is initially an uniform coating of lipids on the PM with density $\rho_0$. An on-state integrin deforms the membrane and thereby induces excess area on it. We argue that this excess area $\delta A = (A-A_0)$ is a measure of the amount of lipid clustering. We estimate it as follows. 
\begin{figure}[H]
%\hspace{4cm} 
\begin{minipage}{0.68\textwidth}
    \includegraphics[width=0.68\linewidth, height=0.10\textheight]{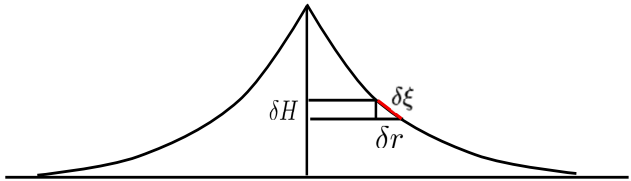}%
        \end{minipage}%
\caption{Schematics of integrin mediated deformation profile. The integrin induced excess area is the measure of the lipid clustering.}
\label{fig_clustering}
\end{figure} The arc-length $\delta \xi$ is related to the increment of the height $H$ and radius $r$ through $ \delta \xi  = \delta r \sqrt{ 1 +(\delta H/\delta r)^2} $ in the limit $\delta r \rightarrow 0$. This leads to  $ 
\delta A  = \int^{2\pi}_{0} d\theta  \int^{R(t)}_{0} r\, dr \,\, [\sqrt{1+ (\partial H/\partial r)^2 } -1]. $ We estimate the clustering of lipids ($\Delta Q$) by calculating the excess area due to the integrin-induced deformation as 
\begin{equation}
\Delta Q(t) = 2 \pi \rho_{0}   \int^{R(t)}_{0} r\, dr \,\, \left(\sqrt{1+ \left(\frac{\partial H}{\partial r}\right)^2 } -1\right).
\label{FlucClustering2D}
\end{equation}
%The second term $-1$ takes care of the diffusive lipids of the membrane projected on a flat surface. 

%%%%%%%%%%%%%%%%%%%%%%%%%%%%%%%%%%%%%%%%%%%%%%%%%%%%%%%%%%%%%%%%%%%%%%%%%%%%%%%%%%%%%%%%%%%%%%%
%%%%%%%%%%%%%%%%%%%%%%%%%%%%%%%%%%%%%%%%%%%%%%%%%%%%%%%%%%%%%%%%%%%%%%%%%%%%%%%%%%%%%%%%%%%%%%%
%%%%%%%%%%%%%%%%%%%%%%%%%%%%%%%%%%%%%%%%%%%%%%%%%%%%%%%%%%%%%%%%%%%%%%%%%%%%%%%%%%%%%%%%%%%%%%%

%%%%%%%%%%%%%%%%%%%%%%%%%%%%%%%%%%%%%%%%%%%%%%%%%%%%%%%%%%%%%%%%%%%%%%%%%%%%%%%%%%%%%%%%%%%%%%%
%%%%%%%%%%%%%%%%%%%%%%%%%%%%%%%%%%%%%%%%%%%%%%%%%%%%%%%%%%%%%%%%%%%%%%%%%%%%%%%%%%%%%%%%%%%%%%%
%%%%%%%%%%%%%%%%%%%%%%%%%%%%%%%%%%%%%%%%%%%%%%%%%%%%%%%%%%%%%%%%%%%%%%%%%%%%%%%%%%%%%%%%%%%%%%%

Assuming circular symmetry around the location of the integrin on a two-dimensional membrane, $R(t)$ in Eq.\,(\ref{FlucClustering2D}) is the separation from the origin where the deformation first becomes zero. 
This separation $R(t)\sim t^{1/4}$ as we discussed before.

\begin{figure}
%\hspace{-1.20cm}
\hspace*{-1.3cm}
\begin{minipage}{0.6 \textwidth}
%\centering
        \includegraphics[width=0.6\linewidth, height=0.2\textheight]{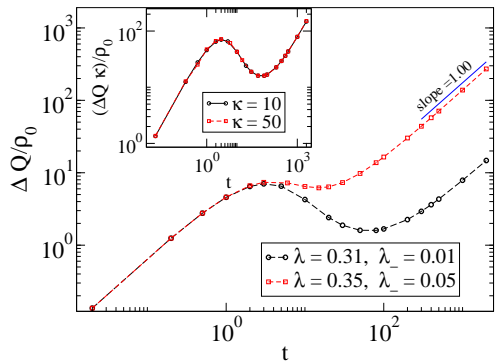}%
        \end{minipage}%
\caption{Time evolution of lipid clustering driven by integrin-induced active region for the tension-less rigid two-dimensional membrane.  This plot is obtained by integrating over full deformation profile and show the clustering for different $\lambda_{-}$. In the inset, we show the evolution of deformation at the position of the integrin for different $\kappa$. We find $\Delta Q \sim \kappa^{-1}$ at all time wherein  we choose $\lambda=0.31\,\text{sec}^{-1}$ and $\lambda_{-}=0.01\,\text{sec}^{-1}$.
}\label{clustering_OFFRate_Kappa}
\end{figure}

The lipid clustering induced by the integrin evolves nonmonotonically in time as shown in Fig.(\ref{clustering_OFFRate_Kappa}). The clustering parameter, $\Delta Q$ grows linearly in both the small and long time limits, and is nonmonotonic in between. 
The initial growth occurs for a time $\sim\tau_{+}$ during the on-state of integrin, while the fall happens during the off-state as the membrane relaxes. 

How does the propensity for clustering depend on the membrane properties? To get an idea, note that clustering is associated with the induced excess area. Since $H(\mathbf{0},t) \sim \kappa^{-1/2}$ and $R(t) \sim \kappa^{-1/4}$, we expect $\Delta Q \sim \kappa^{-1}.$ We confirm this by integrating over the deformation profile for different values of $\kappa$ at different times as shown in the inset in Fig.\,\ref{clustering_OFFRate_Kappa}. 

In \cite{SM2022} we have investigated the effects of going beyond the model studied here in several ways: by allowing integrin to diffuse, by including the effects of surface tension on membrane dynamics and by studying different initial conditions. As argued earlier, the first two points are not likely to matter in the range of parameters pertinent to integrin-membrane dynamics in the focal adhesion region. The third effect, which deals with initial condition variations, shows that the results discussed here, for integrin initially in the on state, are actually representative of a broad set of randomly mixed initial conditions.

\section{III. Discussion and Conclusion}
To summarize, we have constructed a minimal model for lipid clustering driven by a single integrin, through its action on the plasma membrane, as it shuttles stochastically between on and off states.  Within our model, we derived an analytic expression for the deformation, which brings out explicitly its dependence on the microscopic parameters of our model, namely the stochastic on-off dynamics, the bending rigidity of the membrane and the harmonic confinement. The deformation shows an interesting non-monotonic behavior in time. Evidently, our formulation can be readily generalized to deal with the case of many non-interacting integrins by considering the superposition of membrane profiles generated by each, reflecting the linearity of the equations.   On the experimental side, we hope our study would motivate studies to understand single-integrin-induced deformation, the membrane velocity, and associated lipid clustering, while on the theoretical side, it would be interesting to go beyond the independent-integrin assumption by including the effects of integrin clustering.

\section*{ACKNOWLEDGMENT} TS is grateful for the hospitality provided by the S. N. Bose National Centre for Basic Sciences, India. AP acknowledges the support of a research grant under the Prof. T.R. Rajagopalan fund scheme, SASTRA University, India. MB acknowledges support under the DAE Homi Bhabha Chair Professorship of the Department of Atomic Energy, India.

\section*{APPENDIX}
\subsection*{Appendix A: Integrin mediated deformation} Considering the Laplace and Fourier transform of Eq.\,(\ref{FKP_Integrin}), we get     
\begin{eqnarray}
&& -P_{+}(\mathbf{q},0) + (s+\lambda_{+}) \widetilde{P}_{+}(\mathbf{q},s) = \lambda_{-} \widetilde{P}_{-}(\mathbf{q},s) \\
&& -P_{-}(\mathbf{q},0) + (s+\lambda_{-}+D |\mathbf{q}|^2) \widetilde{P}_{-}(\mathbf{q},s) =  \lambda_{+} \widetilde{P}_{+}(\mathbf{q},s), \nonumber \\
\label{IntegrinLaplace}
\end{eqnarray}
 where $P_{+}(\mathbf{q},0)$ is the Fourier transform of the probability of the integrin being in the on-state at time $t=0$. With the initial condition ($t=0$) is that integrin is located at $\mathbf{r}=\mathbf{r}_0$, and the probability of being in the on (off)-state is $p_0$ ($q_0$). We define as $P_{+}(\mathbf{r}, 0)=p_0\,  \delta(\mathbf{r}-\mathbf{r}_0)$ as the probability that an integrin was in on-state, and $P_{-}(\mathbf{r}, 0)= q_0\, \delta(\mathbf{r}-\mathbf{r}_0)$  is the probability that an integrin is in its off-state at location $\mathbf{r}=\mathbf{r}_0$. Evidently, we have $p_0+q_0=1$. Fourier transformation of $P_{+}(\mathbf{r}, 0)$ gives
 \begin{eqnarray}
P_{+}(\mathbf{q}, 0) && = \int d^2 \mathbf{r}\,\,  e^{i \mathbf{q}\cdot \mathbf{r}} \,\, P_{+}(\mathbf{r},0) 
= p_0 \,\, e^{i \mathbf{q} \cdot \mathbf{r}_0} 
\end{eqnarray} and, $ P_{-}(\mathbf{q}, 0) = q_0 \,\,  e^{i \mathbf{q} \cdot \mathbf{r}_0}.$ We substitute $q_0=1-p_0$ in the above equation, and get $ P_{-}(\mathbf{q}, 0) = (1-p_0) \, e^{i \mathbf{q} \cdot \mathbf{r}_0} .$ From Eq.\,(\ref{IntegrinLaplace}), we obtain 
\begin{equation}
\widetilde{P}_{-}(\mathbf{q},s) = \frac{\lambda_{+} \widetilde{P}_{+}(\mathbf{q},s)}{(s+\lambda_{-}+D \mathbf{q}^2)}+ \frac{P_{-}(\mathbf{q},0)}{(s+\lambda_{-}+D \mathbf{q}^2)}     
 \end{equation}
where $\lambda_+$ and  $\lambda_-$ are defined earlier. 
Using the expression for $\widetilde{P}_{-}(\mathbf{q},s)$, we derive
\begin{equation}
\widetilde{P}_{+}(\mathbf{q},s) = \frac{(s+\lambda_{-}+D \mathbf{q}^2)P_{+}(\mathbf{q},0) + \lambda_{-}P_{-}(\mathbf{q},0)}{(s+\lambda_{+})(s+\lambda_{-}+D\mathbf{q}^2)-\lambda_{+}\lambda_{-}} 
\end{equation}
Using $P_{+}(\mathbf{q},0)=p_0 \,\, e^{i \mathbf{q} \cdot \mathbf{r}_0} $ and $P_{-}(\mathbf{q},0)=(1-p_0) \,\, e^{i \mathbf{q} \cdot \mathbf{r}_0} $, and substituting it in the above equation, we obtain
\begin{equation}
\widetilde{P}_{+}(\mathbf{q},s) = \frac{(s + D \mathbf{q}^2)p_{0} +\lambda_{-}}{s^2+(\lambda+D\mathbf{q}^2)s+\lambda_{+}D\mathbf{q}^2} \,\, e^{i \mathbf{q} \mathbf{r}_0}
\label{Laplace_P}
\end{equation}
On the other hand, for a zero-tension membrane, the equation for average deformation can be written as 
\begin{equation}
\frac{\partial H(\mathbf{r},t)}{\partial t} = -\kappa \nabla^4 H(\mathbf{r},t) -\gamma\, H (\mathbf{r},t) + I_0 P_{+}(\mathbf{r},t).      \label{EW_IntegrinBreak}
\end{equation}
Considering the Fourier-Laplace transformation of the above equation, and using the expression for $\widetilde{P}_{+}(\mathbf{q},s)$ from Eq.\,(\ref{Laplace_P}), we get 
\begin{align}
 \widetilde{H}(\mathbf{q},s)=  \frac{I_0 \, (\lambda _- + \left(s+D  \mathbf{q}^2\right) p_0) e^{i \mathbf{q}.\mathbf{r}_0}  }{(s+ \gamma+\kappa \mathbf{q}^4)(s^2 + (\lambda + D \mathbf{q}^2)s+ \lambda_{+}\, D \, \mathbf{q}^2)}.
\label{HtBendingRigidity}
\end{align}
Recalling that $ \mathcal{L}^{-1} \left(\frac{1}{s+\alpha}\right) = e^{-\alpha \, t}$, the average deformation in real time is given by 
\begin{eqnarray}
 H(\mathbf{q},t)  && = I_0 \,\,  p_{+} ( \frac{e^{-(\gamma+\kappa \mathbf{q}^4) t}}{(\lambda-(\gamma+\kappa \mathbf{q}^4))}  -\frac{e^{-\lambda t}}{(\lambda-(\gamma+\kappa \mathbf{q}^4))} )  \nonumber \\ &&+ I_0 \lambda_{-} (\frac{1}{\lambda (\gamma+\kappa \mathbf{q}^4)} - \frac{e^{-\lambda t}}{\lambda ((\gamma+\kappa \mathbf{q}^4) -\lambda)} \nonumber \\ && - \frac{e^{-(\gamma+\kappa \mathbf{q}^4) t}}{(\gamma+\kappa \mathbf{q}^4)(\lambda-(\gamma+\kappa \mathbf{q}^4)) } ) .
\end{eqnarray}
Inverse Fourier transform yields the average deformation in real space: 
\begin{eqnarray}
H(\mathbf{r},t) && =  \int \frac{d^2 q}{[2\pi]^2} \, e^{i \, \mathbf{q}\cdot (\mathbf{r}-\mathbf{r}_0)} ( \, I_0 \, p_{0} \int^{t}_{0} dt'\, e^{-\gamma t'} \, e^{-\lambda (t-t')} e^{-\kappa q^4 t'} \nonumber \\ && + I_0 \, \lambda_{-}   \int^{t}_{0} dt'\, e^{-\kappa q^4 t'} e^{-\gamma t'} \int^{t'}_{0} dt'' \, e^{-(\lambda -\gamma -\kappa q^4)t''} ) 
\label{DeformProf}
\end{eqnarray}
where $\mathbf{r}_0$ is the location of the integrin. In principle, this gives the full deformation profile. %The height at the location of the integrin can excatly be obtained as 
%\begin{eqnarray}
%\nonumber
%&& H(\mathbf{r}_0=\mathbf{0},t)   =\frac{I_0}{\sqrt{64\pi \kappa}} \,\, \left(  p_{0} e^{-\lambda t} \int^{t}_{0} dt'\,\, %%\frac{e^{(\lambda-\gamma) t'}}{\sqrt{t'}}   + \lambda_{-} \sqrt{\frac{\pi}{\gamma}}  \int^{t}_{0}  dt'' e^{-\lambda t''} %\text{Erf}[\sqrt{\gamma(t-t'')}]  \right) \end{eqnarray}

\subsection*{Appendix B: Velocity} The rate of the mean deformation defines the mean velocity along the normal to the membrane.
We find, from Eq.\,(\ref{DeformProf}), the profile $V(\mathbf{r},t)$ as  
\begin{eqnarray}
V(\mathbf{r},t) && = I_0\, e^{-\gamma t}  \int \frac{d^2 q}{[2\pi]^2} e^{i \mathbf{q} \cdot (\mathbf{r}-\mathbf{r}_0)} [ p_{0} e^{-\kappa q^4 t}  \nonumber \\ && -p_{0} \,\lambda \, e^{ (\gamma -\lambda)t} \int^{t}_{0} dt' e^{-(\gamma -\lambda)t'} e^{-\kappa q^4 t'} \nonumber \\ && + \lambda_{-} \,\, e^{-\kappa q^4 t} \int^{t}_{0} dt' e^{-(\lambda-\gamma-\kappa q^4)t'} ] 
\label{VelProfile}
\end{eqnarray}
On substituting $\mathbf{r}=\mathbf{r}_0=0$, we obtain a closed form expression for velocity :  
\begin{eqnarray}
V(\mathbf{0},t) = \frac{I_0 e^{-\gamma t}}{\sqrt{64 \pi \kappa}} [ \frac{p_{0}}{\sqrt{t}} - &&\lambda \left(p_{0} -\frac{\lambda_{-}}{\lambda}\right) \sqrt{\pi} e^{-(\lambda-\gamma) t} \nonumber \\ && \frac{\text{Erf}[\sqrt{(\gamma -\lambda)t}]}{\sqrt{(\gamma -\lambda)}} ]  
\end{eqnarray}

\end{document}